\documentstyle[prd,aps,epsfig,preprint,tighten]{revtex}
\begin{document}
\preprint{                                       BARI-TH/326-98, hep-ex/9901012}
\title{        Test of non-standard neutrino properties with 
                     the BOREXINO source experiments
}
\author{A.\ Ianni$^{a,b}$~\footnote{email: ianni@lngs.infn.it}, 
        D.\ Montanino$^{c}$~\footnote{email: montanin@sissa.it}, and 
        G.\ Scioscia$^{d}$~\footnote{email: scioscia@ba.infn.it}}
\address{
$^{a}$Dipartimento di Fisica dell'Universit\`a dell'Aquila,\\
     Localit\`a Coppito, I-67010 L'Aquila, Italy\\
$^{b}$Laboratori Nazionali del Gran Sasso,\\
     S.S. 17bis Km 18+910, I-67010 Assergi (L'Aquila)\\
$^{c}$Scuola Internazionale Superiore di Studi Avanzati (S.I.S.S.A.),\\ 
      Via Beirut 4, I-34014 Trieste, Italy \\
$^{d}$Dipartimento di Fisica dell'Universit\`a di Bari,\\
      Via Amendola 173, I-70126 Bari, Italy
}\maketitle
\begin{abstract}
We calculate the event rates induced by  high-intensity radioactive sources of
$\nu_{e}$ ($^{51}$Cr) and of $\bar{\nu}_{e}$ ($^{90}$Sr), to be located near
the BOREXINO detector. Calculations are performed both in the standard case and
assuming non-standard properties of neutrinos, including flavor oscillations,
neutrino electromagnetic interactions, and deviations from the standard vector
and axial couplings in the $\nu_{e}$-$e$ interaction. It is shown that, in some
cases, the current limits on non-standard neutrino properties can be
significantly improved.
\end{abstract}
\pacs{\\ PACS number(s):   13.15.+g;  14.60.Pq;  14.60.St;  25.30.Pt;  95.55.Vj}

                      \section{Introduction}

	The BOREXINO experiment \cite{Borexino1,Borexino2,URL}, under
construction at Gran Sasso, is designed to study $^{7}$Be solar neutrinos
\cite{Rag} through a real-time, low-background detector, consisting of a nylon
sphere (8.5 m in diameter) filled with high purity organic scintillator
(pseudocumene, C$_{9}$H$_{12}$).

	The apparatus can be calibrated through external (anti-)neutrino
sources as well as through light sources \cite{Light}. For the first type of
calibration experiment, two sources have been considered: $^{51}$Cr and
$^{90}$Sr. The $^{51}$Cr source generates $\nu_{e}$ through the reaction
$^{51}{\rm Cr}+e^-\rightarrow\;^{51}{\rm V}+\nu_{e}$, with a half-life 
$\tau^{\rm Cr}_{1/2}$ of 27.7~days and four energy lines:  $E_{1}=0.751$~MeV
(9\%), $E_{2}=0.746$~MeV (81\%), $E_{3}=0.431$~MeV (1\%), $E_{4}=0.426$~MeV
(9\%).

\begin{figure}
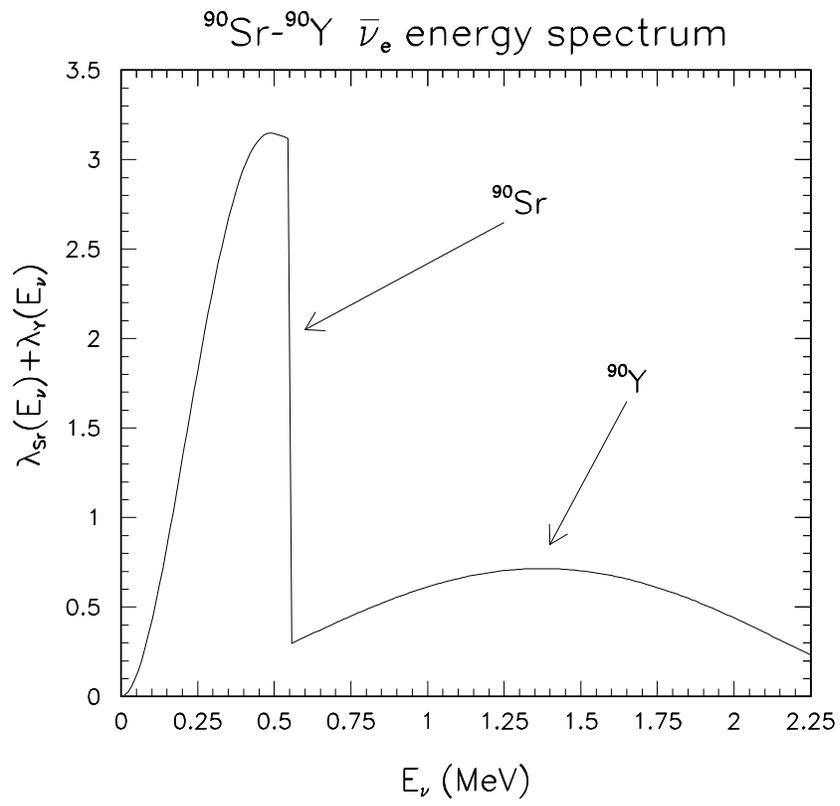

\caption{Spectrum of the $^{90}$Sr antineutrino source.}
\label{spectrum}
\end{figure}

	The $^{90}$Sr source generates $\bar{\nu}_{e}$ through the reaction
$^{90}{\rm Sr} \rightarrow\;^{90}{\rm Y}+ \bar{\nu}_{e}+e^-$ ($\tau^{\rm
Sr}_{1/2}\sim 28$~y) followed by $^{90}{\rm Y} \rightarrow\;^{90}{\rm Zr}+
\bar{\nu}_{e}+e^-$ ($\tau^{\rm Y}_{1/2}\sim 64.8$~h). Since $\tau^{\rm
Y}_{1/2}\ll \tau^{\rm Sr}_{1/2}$, one can simply assume that two
$\bar{\nu}_{e}$'s are produced for each $^{90}$Sr nucleus decay. The total
standard spectrum of this source is given by $\lambda(E_{\nu})=\lambda_{\rm
Sr}(E_{\nu})+\lambda_{\rm Y}(E_{\nu})$, where each $\lambda_{i}$ ($i\in\{{\rm
Sr},{\rm Y}\}$) is calculated using Fermi theory \cite{Kono}:
\begin{equation}  
\lambda_{i} \left(E_{\nu}\right)=
A_{i}\frac{x_{i}}{1-e^{-x_{i}}}\left(Q_{i}+m_{e}-E_{\nu}\right) E_{\nu}^{2}
\sqrt{\left(Q_{i}+m_{e}-E_{\nu}\right)^{2}-m_{e}^{2}} 
\, .\label{lambda} \end{equation}
In Eq.~(\ref{lambda}), $A_{i}$ is a normalization  factor [so that $\int
dE_{\nu}\lambda_i(E_{\nu})=1$], $Q_{i}$  is the endpoint energy ($Q_{\rm
Sr}=0.546$~MeV and $Q_{\rm Y}=2.27$~MeV), $m_{e}$ is the electron mass, and
\begin{equation}
x_{i}=2\pi Z_i\alpha_{\rm e.m.} \frac{Q_{i}+m_{e}-E_{\nu}}
{\sqrt{\left(Q_{i}+m_{e}-E_{\nu}\right)^{2}-m_{e}^{2}}} \,
,\label{lambda2}\end{equation}
$Z_{i}$ being the atomic number of the decaying nuclei and $\alpha_{\rm
e.m.}$=1/137.036 \cite{PDB}. The standard $^{90}$Sr-$^{90}$Y spectrum is shown
in Fig.~\ref{spectrum}.

	Artificial neutrino sources of known activity and spectra can be used
to probe nonstandard neutrino properties, such as flavor oscillations or
magnetic moment \cite{Borexino3,Fiore,Ianni} or non-standard neutrino couplings
(for example through the exchange of an additional $Z$ boson \cite{Valle}). In
particular, in \cite{Sinev} the combined analysis of $\bar{\nu}_{e}$-$e$
scattering and inverse beta decay using only a $^{90}$Sr source has been
proposed for studying both oscillation and non-standard neutrino couplings.

	In this paper we study in detail how to use the $^{51}$Cr and $^{90}$Sr
sources to probe flavor oscillations, electromagnetic properties, and
deviations of the axial and vector couplings $g^{\nu e}_{V}$ and  $g^{\nu
e}_{A}$ from their standard value in $\nu_{e}$-$e$ scattering. The sensitivity
of the experiments depends on the initial activity of the sources, which is
constrained by technical and economical budget limits.~\footnote{
A high intensity source with a proper shielding costs about 1M\$/MCi
\cite{Barba}.} 
As a reasonable reference value we use 5~MCi$=1.85\times 10^{17}$~decay/s for
each source activity.

	The plan of this work is the following: in Sec.~II we describe the
calibration experiments and calculate the standard expectations. In Sec.~III we
calculate the effect of oscillations into active or sterile states and
determine the BOREXINO sensitivity to the neutrino mass and mixing. In Sec.~IV 
we study how BOREXINO can probe neutrino magnetic and  anapole moments. In
Sec.~V we show how this experiment can put interesting limits on the
$\nu_{e}$-$e$ vector  and axial couplings. Finally, we draw our conclusions in
Sec.~VI.

    \section{Description of the experiment and standard expectations}

	In this section we describe the radioactive source experiments and the
standard rates induced by the sources. We assume the following technical setup:
100 tons ($6\times 10^{30}$ protons, $3.3\times 10^{31}$ electrons) of
spherical ($R=3$~m) fiducial volume (FV) and 5~MCi pointlike sources placed 
at distance $D=8.25$~m from the detector center.

	In the $^{51}$Cr source experiment ($\nu_{e}$) the detection process is
neutrino-electron elastic scattering, while in the $^{90}$Sr experiment
($\bar{\nu}_{e}$) the interactions are elastic scattering and inverse
$\beta$-decay. The scattering events are identified through the scintillation
light from the electron. The inverse $\beta$-decays are identified through the
delayed coincidence between the prompt positron annihilation signal and the
neutron capture $\gamma$'s.

	The standard (std) differential cross section of the scattering process
$\nu_{e,\mu}+e\rightarrow\nu_{e,\mu}+e$, as a function of the the incident
neutrino energy $E_{\nu}$ and of the recoil electron kinetic energy $T_{e}$, is
\cite{tHooft}:
\begin{equation}
\frac{d\sigma^{\rm std}_{e,\mu}(E_{\nu},T_{e})}{dT_{e}}=
\frac{G_{F}^{2}m_{e}}{2\pi }\left[\left(C_{V}+C_{A}\right)^{2}+
\left(C_{V}-C_{A}\right)^{2}\left(1-\frac{T_{e}}{E_{\nu }}\right)^{2}+
\left(C_{A}^{2}-C_{V}^{2}\right)\frac{m_{e}T_{e}}{E_{\nu}^{2}}\right]
\, ,\label{diffcross}\end{equation}
where $C_{V,A}=g^{\nu e,{\rm std}}_{V,A}+\delta$ and $g^{\nu e,{\rm
std}}_{V}=2\sin^{2}\theta_{W}-\frac{1}{2}$, $g^{\nu e,{\rm
std}}_{A}=-\frac{1}{2}$, with $\delta=1$ ($\delta=0$) if the incident neutrino
is a $\nu_{e}$ ($\nu_{\mu}$). For antineutrinos, $C_{A}\to -C_{A}$. We take
$\sin^{2}\theta_{W}=0.2312$ \cite{PDB}.

	The cross section in a definite energy window $T_{e}\in
[T_{e,1},T_{e,2}]$ is given by
\begin{equation}
\sigma_{e,\mu}(E_{\nu})=\int_{0}^{E_{\nu}/(1+m_{e}/2E_{\nu})}dT_{e}\,W(T_{e}) 
\frac{d\sigma^{\rm std}_{e,\mu}(E_{\nu},T_{e})}{dT_{e}}
\, ,\label{crossec}\end{equation}
where $W(T_{e})$ accounts for the finite detector resolution \cite{Faid},
\begin{equation}
W(T_{e})=\frac{1}{2}
\left[{\rm Erf}\left(\frac{T_{e,2}-T_{e}}{\sqrt{2}\sigma_{T_{e}}}\right)-
      {\rm Erf}\left(\frac{T_{e,1}-T_{e}}{\sqrt{2}\sigma_{T_{e}}}\right)\right]
\, .\label{W}\end{equation}
We assume $\sigma_{T_{e}}/{\rm keV}=48\sqrt{T_{e}/{\rm MeV}}$, as obtained by
the MonteCarlo simulations of the apparatus \cite{Giam}.

	Our reference choice for the energy window of the $\nu_{e}$-$e$
scattering experiment is $T_{e}\in[0.25,0.7]$~MeV. The lower limit efficiently
cuts the $^{14}$C decay background, and the upper limit is safely above the
Compton edge ($T_{e,{\rm max}}=0.56$~MeV) for the electrons scattered by
$^{51}$Cr neutrinos. For the  $\bar{\nu}_{e}$-$e$ scattering we choose
$T_{e}\in[0.25,1]$~MeV, the upper limit now being determined by the $^{40}$K
contaminant [which emits $\gamma$'s (B.R.=10\%, $E_{\gamma}=1.460$~MeV) and
$\beta$'s (B.R.=90\%, $T_{e}\leq 1.32$~MeV)] in the scintillator.

	For the inverse $\beta$-decay process $\bar{\nu}_{e}+p\rightarrow
e^{+}+n$, (characterized by a threshold $E_{\nu,{\rm min}}=1.804$~MeV), the
cross section is \cite{Yama}
\begin{equation}
\sigma_{e}(E_{\nu})=\sigma_{0}\left(E_{\nu}-Q\right) 
\sqrt{\left(E_{\nu}-Q\right)^{2}-m_{e}^{2}}
\, ,\label{invb}\end{equation}
with $\sigma_{0}=94.55\times 10^{-45}$~cm$^{2}/{\rm MeV}^{2}$, and
$Q=1.2933$~MeV. We actually improve Eq.~(\ref{invb}) to account for the weak
magnetism and bremsstrahlung corrections, as described in \cite{Correct}.

	For an exposure time $t_{\rm ex}$, the expected number of events,
$N_{0}$, is given by
\begin{equation}
N_{0}=n_{t}\langle\sigma_{e}\rangle \times
\int_{t_{\rm tr}}^{t_{\rm ex}+t_{\rm tr}} dt'\, I(t') \times 
\int_{\rm FV} \frac{d^{3} x}{4\pi\delta^{2}_{x}}
\, ,\label{n0}\end{equation}
where $n_{t}$ is the volume density of targets in the fiducial volume FV,
$t_{\rm tr}$ is the ``transport'' time elapsed from the source activation to
the beginning of the source experiment, $I(t)=I_{0}\exp(-t/\tau)$ is the
intensity of the source ($I_{0}=5$~MCi), $\delta_{x}$ is the distance between 
the source and the generic point $x$ in the detector volume, and
\begin{equation}
\langle\sigma_{e}\rangle=\int dE_{\nu}\, \lambda(E_{\nu})\sigma_{e}(E_{\nu})
\, .\label{sigmaave}\end{equation}
Given the geometry of the experiment, the integral over the volume can be
performed analytically and Eq.~(\ref{n0}) can be recast in the following,
compact form:
\begin{equation}
N_{0}=N_{t}\Phi_{0}F(R/D)\langle\sigma_{e}\rangle\Gamma(t_{\rm ex},t_{\rm tr})
\, ,\label{n01}\end{equation}
where $N_{t}=n_{t}\times V$, $\Phi_{0}=I_{0}/(4\pi D^{2})$, and  the function
$F$ is given by
\begin{equation}
F(h)=\frac{3}{2h^{3}}\left[h-\frac{1-h^{2}}{2}
\ln\left(\frac{1+h}{1-h}\right)\right]
\label{F}\end{equation}
[where in our case, $F(R/D)=1.028$] and $\Gamma(t_{\rm ex} ,t_{\rm
tr})=\tau\exp(-t_{\rm tr}/\tau)\times [1-\exp(-t_{\rm ex}/\tau)]$. We assume
$t_{\rm tr}=5$~days for both sources (Cr and Sr). For the $^{51}$Cr experiment,
we take $t_{\rm ex}=60$~days, which maximizes the signal-to-noise ratio, as
shown in \cite{Ianni}. For the $^{90}$Sr experiment, the useful time limit can
be determined by the condition that the statistical uncertainty of the rate
reaches the size of the systematic error of the source activity (about 1\%)
\cite{uncert}. This leads us to $t_{\rm ex}=1/2$~years as a realistic exposure
time. (A longer experiment would also interfere with the measurement of the
solar $\nu$ rate, which is the main goal of BOREXINO.)

	Now we discuss the background rates. For the scattering events, the
background events are due both to solar neutrinos interactions and to the
internal decays of radiocontaminants. The background rate $R_{B}$ can be
measured during the ``source off'' operation of the apparatus, and the number
of background events is then simply given by $N_{B}=R_{B}\times t_{\rm ex}$.
For Standard Solar Model expectations \cite{Bahcall}, the total background rate
is expected to be $R_{B}=73$~events/day for events with $T_{e}\in
[0.25,07]$~MeV and $R_{B}=97$~events/day when $T_{e}\in [0.25,1]$~MeV
\cite{Ianni}. For the inverse $\beta$-decay events, the delayed coincidence
signature allows an almost complete background rejection (apart from
10~event/year due to antineutrinos coming from nuclear reactors \cite{Stefan}),
so we simply set $R_{B}\simeq 0$ in this case.

	Although the background can be measured in the source off mode and
subtracted from the total signal, it contributes to the uncertainties through
statistical fluctuations. The total (signal+background) $1\sigma$ uncertainty
of the signal is
\begin{equation}
\delta_{N_{0}} =\sqrt{N_{B}+N_{0}\left(1+\delta_{A}^{2}N_{0}\right)}
\, ,\label{deltaN}\end{equation}
where $\delta_{A}=0.1$ is the estimated uncertainty of the source activity and
errors have been added in quadrature. A deviation from the standard
expectations can be evidenced at 90\%~C.L.\ if the measured rate $N$ satisfies
\begin{equation}
\left|\frac{N}{N_{0}}-1\right| \geq \varepsilon_{90}
\, ,\label{Limit}\end{equation}
where $\varepsilon_{90}=2.146\times\delta_{N_{0}}/N_{0}$ (for the two degrees
of freedom that we will consider).

	Finally, in this paper we compare the BOREXINO performance with MUNU
\cite{MUNU}, a reactor experiment mounted at the Bugey laboratory, designed
explicitly for studying $\bar{\nu}_{e}$-$e$ scattering. MUNU has been running
since August 1998. For this experiment we consider the energy window
$[T_{e,1},T_{e,2}]=[0.5,1]$~MeV \cite{MUNUexp}. The energy resolution is
$\sigma_{T_{e}}/{\rm keV}=140\sqrt{T_{e}/{\rm MeV}}$ \cite{Jonkmans}. The
energy spectrum for the $\bar{\nu}_{e}$ was taken from \cite{MUNUspect}. The
standard expected rate is 5.3~events/day while the background is estimated in 
6~events/day \cite{MUNUexp}. The uncertainty on the neutrino flux from the
reactor is $\sim$5\% \cite{MUNUexp}. As reference, we set $t_{\rm ex}=1$~y.

	In Tab.~\ref{tab1} we report the background, the standard expectation, 
the uncertainty, and the $\varepsilon_{90}$ for the three measure ($\nu$-$e$
and $\bar{\nu}$-$e$ scattering, and inverse $\beta$-decay) of interest for this
paper and for the MUNU experiment.

           \section{Probing flavor oscillations}

	Neutrino flavor oscillations \cite{Pont} represent a viable solution to
the so-called solar neutrino problem \cite{Ba89} and to the atmospheric
neutrino anomaly \cite{Atm}. This phenomenon can also be probed at
accelerators  \cite{Bo92,KARM,CDHSW,At96,Us86,Ar98} and reactors
\cite{Zacek,Ac95,Vi94,Ap98}. The flavor survival probability of a neutrino with
energy $E_{\nu}$, at a distance $L$ from the source is
\begin{equation}
P(E_{\nu},L)=1-\frac{1}{2}\sin^{2}2\theta 
\left[1-\cos\left(\frac{\delta m^{2}}{2E_{\nu}}L\right)\right]
\, ,\label{Pveve}\end{equation}
where $\theta$ is the mixing angle and $\delta m^{2}=m_{2}^{2}-m_{1}^{2}$ is
the difference between the square of the two neutrino masses. (We have assumed
only two neutrino families for simplicity).

	A deficit of the measured rate in the BOREXINO source experiments might
signal neutrino oscillations, i.e., the disappearance of the initial flavor
$\nu_{e}$ into either active states (say, $\nu_{\mu}$) or sterile states
($\nu_s$). We assume two-family $\nu_{e}\rightarrow\nu_{\mu}$ or $\nu_{e}
\rightarrow\nu_{s}$ oscillations. In the presence of oscillations,
Eq.~(\ref{n0}) transforms in the following way:
$$
N(\delta m^{2},\sin^{2} 2\theta)=n_{t}\int_{t_{\rm tr}}^{t_{\rm ex}+t_{\rm tr}} 
dt'\, I(t')\int dE_{\nu}\, \lambda(E_{\nu})
$$
\begin{equation}
\times \int_{\rm FV}\frac{d^{3} x}{4\pi\delta^{2}_{x}}
\left[\phantom{\frac{}{}}\!
\sigma_{e}(E_{\nu})P(E_{\nu},\delta_{x})+
\sigma_{\rm NC}(E_{\nu})\left(1-P(E_{\nu},\delta_{x})\right)
\phantom{\frac{}{}}\!\!\right]
\, .\label{nP}\end{equation}
In Eq.~(\ref{nP}), $\sigma_{\rm NC}=\sigma_{\mu}$ for
$\nu_{e}\rightarrow\nu_{\mu}$ and $\sigma_{\rm NC}=0$ for
$\nu_{e}\rightarrow\nu_{s}$ and inverse beta-decay. Inserting the expression
for the probability~(\ref{Pveve}) in Eq.~(\ref{nP}), we obtain, through 
Eq.~(\ref{n01}):
\begin{equation}
N(\delta m^{2},\sin^{2} 2\theta)=
N_{0}\left[1-\frac{1}{2}\sin^{2}2\theta 
\left(1-\rho-\gamma \left(\delta m^{2}\right)\right)\right]
\, ,\label{nP1}\end{equation}
where $N_{0}$ is the standard expectation and
\begin{eqnarray}
\rho =\frac{\int dE_{\nu}\, \sigma_{\rm NC}(E_{\nu}) \lambda(E_{\nu})}
           {\int dE_{\nu}\, \sigma_{e}(E_{\nu}) \lambda(E_{\nu})}=
	   \left\{
	   \begin{array}{ll}
		0.222	& \text{for $\nu$-$e$ scattering}, \\
		0.433	& \text{for $\bar{\nu}$-$e$ scattering}, \\
		0	& \text{for $\nu_{e}\rightarrow\nu_{s}$
			         and inverse $\beta$ decay}, \\
	   \end{array} 
	   \right.
\label{rho}\end{eqnarray}
and
\begin{equation}
\gamma(\delta m^{2})=
\frac{\int dE_{\nu}\,\left[\sigma_{e}(E_{\nu})-\sigma_{\rm NC}(E_{\nu})\right] 
    \lambda(E_{\nu})g(R/D,\delta m^{2}D/2 E_{\nu})}
     {\int dE_{\nu}\,\sigma_{e}(E_{\nu})\lambda(E_{\nu})}
\, ,\label{gamma}\end{equation}
where the function $g$ is given by
\begin{equation}
g=\frac{1}{F(R/D)}\frac{3D^{2}}{R^{3}}\int_{\rm FV}
\frac{d^{3}x}{4\pi \delta_{x}^{2}}\cos\frac{\delta m^{2}\delta_{x}}{2E_{\nu}}
\, .\label{gdraft}\end{equation}
This function can be calculated analytically (see the appendix).

\begin{figure}
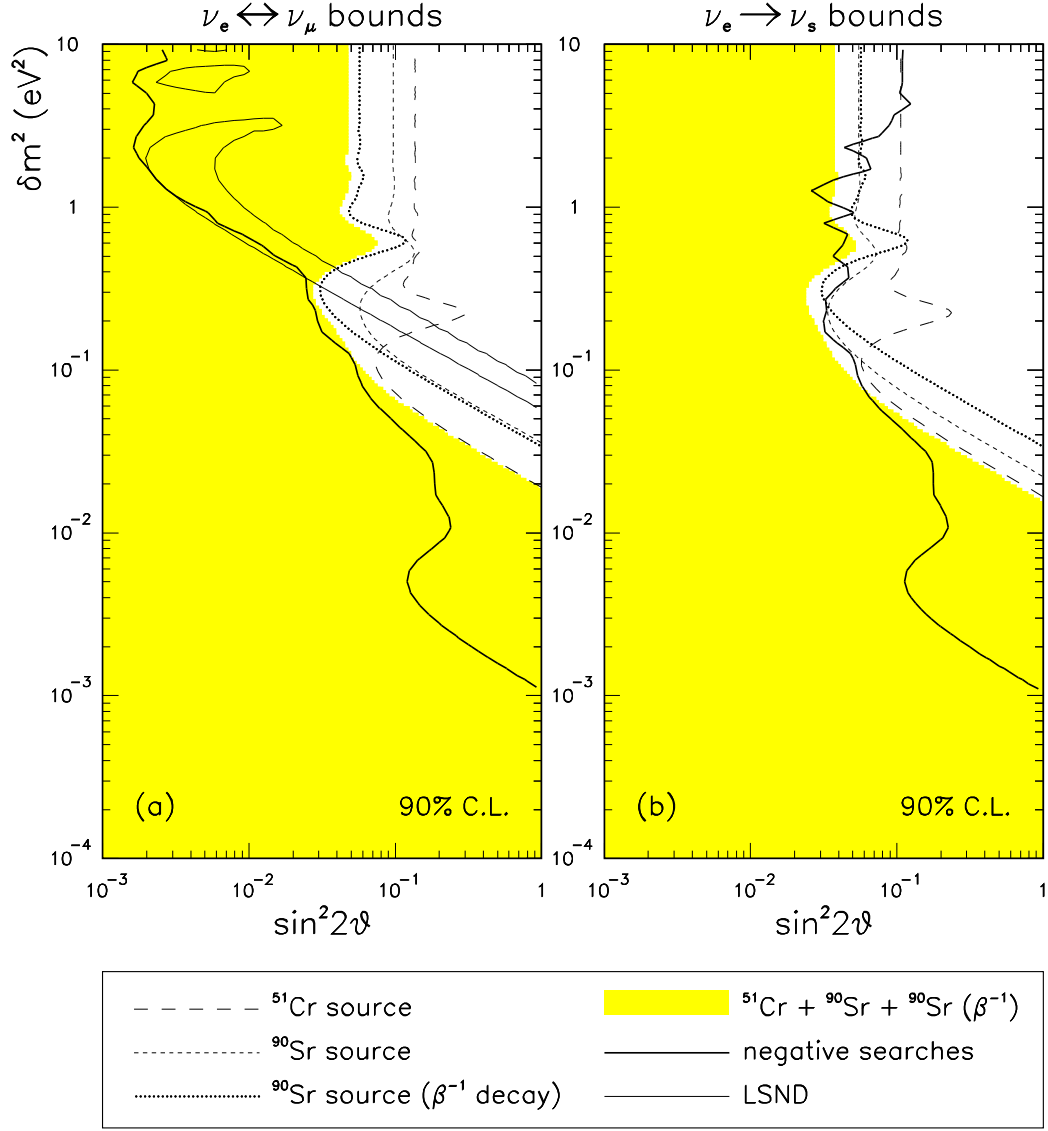

\caption{Prospective 90\% C.L.\ sensitivity contours in the oscillation 
	 parameters plane for the BOREXINO source experiments in the case of 
	 $\nu_{e} \rightarrow\nu_{\mu}$ and $\nu_{e}\rightarrow\nu_{s}$ 
	 transitions [panel (a) and (b), respectively]. Short-dashed line: 
	 $\nu$-$e$ scattering of the $^{90}$Sr antineutrinos; long-dashed line: 
	 $\nu$-$e$ scattering of the $^{51}$Cr neutrinos; dotted line: inverse 
	 $\beta$-decay; gray area: combined Sr+Cr experiments. The 90\% C.L.\ 
	 limits coming from negative evidences of oscillations (solid thick 
	 line) and the LSND allowed area (thin solid line) are also shown.}
\label{osc}
\end{figure}

	From Eqs.~(\ref{nP1}) and~(\ref{Limit}) we obtain a compact form for
the 90\% limit in the plane $(\delta m^{2},\sin^{2} 2\theta)$:
\begin{equation}
\sin^{2}2\theta 
\left(1-\rho-\gamma\left(\delta m^{2}\right)\right)=2\varepsilon_{90}
\, ,\label{osclim}\end{equation}
In Fig.~\ref{osc} we show the 90\% C.L.\  contours in the plane $(\delta m^{2},
\sin^{2} 2\theta)$ for the $^{90}$Sr antineutrinos in the case of $\nu$-$e$
scattering (short-dashed line) and inverse $\beta$-decay (dotted line), and the
for $^{51}$Cr neutrinos (long-dashed line), for the
$\nu_{e}\rightarrow\nu_{\mu}$ [panel~(a)] and $\nu_{e}\rightarrow\nu_{s}$
[panel~(b)] transitions. The gray area is the combined fit (i.e., the  allowed
zone if no deficit were found). Superposed we show also the 90\% C.L.\   bounds
coming from negative evidence for oscillations (solid thick line) and  the LSND
allowed area (solid thin line).

	The function $\gamma(\delta m^{2})$ drops rapidly to zero for $\delta
m^{2}>1$~eV$^{2}$, because of the rapid oscillation of the cosine term in
Eq.~(\ref{gdraft}). Consequently, for $\delta m^{2}>1$~eV$^{2}$, the testable
value of $\sin^{2} 2\theta$ tends to the constant value $2\varepsilon_{90}
/(1-\rho)$. As expected, in the case of $\nu$-$e$ and $\bar{\nu}$-$e$
scattering, the sensitivity in the $\nu_{e}\rightarrow\nu_{s}$ channel is thus
better, due to the absence of the $\rho$ term. The higher statistics attainable
make the $^{90}$Sr source more appropriate to check lower values of $\sin^{2}
2\theta$ for high values of $\delta m^{2}$. On the other hand, the low energy
of the main decay branch of the $^{51}$Cr source make it more appropriate to
check lower values of $\delta m^{2}$, although it cannot compete with the CHOOZ
sensitivity.

	In Fig.~\ref{osc}(a) and (b), the thick solid lines represent the 90\%
C.L.\ exclusion contours due to the $\chi^{2}$-combination of the negative
results coming from short baseline reactor experiments (Bugey \cite{Ac95},
Krasnoyarsk \cite{Vi94}, G{\"osgen} \cite{Zacek}) and from the long baseline
reactor experiment CHOOZ \cite{Ap98}, searching for the disappearance channel
($\bar{\nu}_{e} \leftrightarrow \bar{\nu}_{e}\/$), together with the negative
results from the accelerator experiments E776 ($\nu_{\mu}\leftrightarrow
\nu_{e}$) \cite{Bo92}, KARMEN2 ($\nu_{\mu}\leftrightarrow\nu_{e}$) \cite{KARM},
and CDHSW ($\nu_{\mu}\leftrightarrow \nu_{\mu}$) \cite{CDHSW}. In  the  case of
$\nu_{e}\rightarrow\nu_{s}$ oscillations [panel(b)] only the  disappearance
channel can be probed (through reactor experiments).

	In the panel (a), the thin solid line defines the 90\% C.L.\ favored
region delimited by the positive signal of the LSND experiment \cite{At96} as
obtained in \cite{Fo97}. Notice that only a very small region of the parameter
space  allowed by LSND survives the comparison with negative searches.

	From panel~(a) of Fig.~\ref{osc} we see that the zone tested by the
BOREXINO source experiments, (in the hypothesis of $\nu_{e}\rightarrow
\nu_{\mu}$ oscillations) is already contained in the current bounds. Therefore,
if a significative difference from the standard expectation is  found, it
cannot be interpreted as $\nu_{e}\rightarrow\nu_{\mu}$ oscillations.

	However, in the case of $\nu_{e}\rightarrow\nu_{s}$ oscillations
[panel~(b)] the $^{90}$Sr source can improve the current bounds for $\delta
m^{2}>1$~eV$^{2}$ and in the region around $(\delta m^{2},\sin^{2}
2\theta)\sim (3\times 10^{-1}$~eV$^{2},3\times 10^{-2})$. The higher
sensitivity of BOREXINO for $\delta m^{2}\to\infty$ with respect the reactor
experiments is mainly due to the higher statistics and the lower background. In
particular, the bound on $\sin^{2} 2\theta$ for $\delta m^{2}\geq 3$~eV$^{2}$
can be shifted to $5\times 10^{-2}$ ($4\times 10^{-2}$ if the combined fit is
considered), thus improving the existing limits by about a factor 2.
Moreover,since both $\bar{\nu}_{e}$ scattering and inverse $\beta$-decay can
probe the same values of $\sin^{2} 2\theta$  for $\delta m^{2}>1$eV$^{2}$,
cross checks are possible. If a deficit in the counting is found in both cases,
this could be interpreted as a signal for $\nu_{e}\rightarrow\nu_{s}$
transition.

	Finally, the $^{51}$Cr experiment can strongly improve the present
bounds on $\nu_{e}$ transitions fixed by the calibration source experiment in
GALLEX \cite{BKL}. In particular, in the case of $\nu_{e}\rightarrow\nu_{\mu}$
oscillations, the bound on $\delta m^{2}$ would be lowered by a factor 10, from
$\sim 10^{-1}$~eV$^{2}$ to $\sim 10^{-2}$~eV$^{2}$. (The lower bound on
$\sin^{2} 2\theta$ is now fixed by KARMEN to 0.052, stronger than the BOREXINO
one.) In the case of $\nu_{e}\rightarrow\nu_{s}$ transition, BOREXINO would fix
also the bound on $\sin^{2} 2\theta$ to 0.1 --- two times better than GALLEX.
If an oscillation signal were find in the $\nu_{e}$ channel but not in the
$\bar{\nu}_{e}$ channel this would be an evidence for CP violation.

       \section{Implications for neutrino e.m.\ form factors}

	Neutrinos can interact with photons through a possible magnetic dipole
moment or anapole moment (also called charge radius). The effective
$\nu$-$\gamma$ interaction Lagrangian is \cite{Vogel,Salati}
\begin{equation}
{\cal L}^{\rm e.m.}_{\nu}=\frac{\langle r_{\nu}^{2}\rangle}{6}
\bar{\psi}_{\nu}\gamma_{\alpha}\psi_{\nu}\Box A^{\alpha}-
\frac{\mu_{\nu}}{4 m_{e}}
\bar{\psi}_{\nu}\sigma_{\alpha\beta}\psi_{\nu}F^{\alpha\beta}
\, ,\label{emlagr}\end{equation}
where $\psi_{\nu}$ is the neutrino field, $\langle r_{\nu}^{2}\rangle$ is the
anapole moment and $\mu_{\nu}$ the neutrino magnetic moment. The mere existence
of a neutrino Dirac mass implies an effective neutrino magnetic moment equal to
$\mu_{\nu}=3.2\times 10^{-19} \mu_{B}(m_{\nu}/{\rm eV})$ \cite{Lee}, where
$\mu_{B}=e/2m_{e}$ is the Bohr magneton. Regarding the anapole moment, the
situation is controversial. Some authors assert that the  effective anapole
moment coming from the radiative corrections of the neutrino  vertex in the
Standard Model is not gauge invariant and then cannot be a physical observable
\cite{ana0}, while in \cite{Degrassi} it is claimed that a gauge invariant part
can be extracted, yielding a value $\langle r_{\nu_{e}}^{2} \rangle_{\rm
std}\simeq 0.4\times 10^{-32}$~cm$^{2}$ for a top quark mass of 175~GeV.
Conservatively one can say that values of this two e.m.\ form factors larger
than quoted above would be an indication for non-standard neutrino physics.

	Stringent bounds on the neutrino e.m.\ form factors come from
astrophysical  arguments. For example, if the neutrino anapole moment exceeds
$\sim 7\times 10^{-32}$~cm$^{2}$, escaping neutrinos would overcool stars and
hence should modify the color-magnitude diagram of globular clusters
\cite{Salati}. Moreover, the energy loss in red giants in globular clusters via
the plasmon decay $\gamma^{*}\to \nu_{R}\bar{\nu}_R$ mediated by neutrino
magnetic moment would be too large \cite{Raff} unless $\mu_{\nu}\leq 2\times
10^{-12}\mu_{B}$. A rather stringent limit comes also from the SN1987A,
$\mu_{\nu}\leq 5\times 10^{-13}\mu_{B}$ \cite{Supernova}.

	However, the only direct experimental constraint on the $\nu_{e}$
magnetic moment ($\mu_{\nu_{e}}<1.8\times 10^{-10}\mu_{B}$), comes from reactor
experiments sensitive to $\bar{\nu}_{e}$-$e^{-}$ elastic scattering
\cite{Derbin}. As regard the electron neutrino anapole moment $\langle
r_{\nu_{e}}^{2}\rangle$, the more stringent limit comes from the LAMPF
Collaboration which quotes $-7.6\leq\langle r_{\nu_{e}}^{2} \rangle/10^{-32}
{\rm cm^{2}} \leq 10.5$ \cite{LAMPF,Salati}. Improved bounds are expected from
MUNU \cite{MUNU}. In the following, we will also discuss this experiment in
comparison with BOREXINO.

	The possibility to search for a neutrino magnetic moment using an
external neutrino source was addressed by the BOREXINO Collaboration in 1991 
\cite{Borexino2} and then studied in \cite{Fiore} and \cite{Ianni}. In this
section we study $\nu$-$e$ scattering process in the general case, i.e., with
non-zero magnetic and anapole moments.

	From the Lagrangian in Eq.~(\ref{emlagr}), the $\nu_{e}$-$e$
differential cross section is obtained \cite{Vogel,Salati}:
$$
\frac{d\sigma(E_{\nu},T_{e})}{dT_{e}}=
\frac{d\sigma^{\rm std}(E_{\nu},T_{e})}{dT_{e}}+
\frac{\pi\alpha^{2}_{\rm e.m.}\mu_{\nu}^{2}}{m_{e}^{2}}
\left(\frac{1}{T_{e}}-\frac{1}{E_{\nu}}\right)
$$$$
+\langle r_{\nu}^{2}\rangle\frac{\sqrt{2}G_{F}m_{e}}{3}
\left[\left(C_{V}+C_{A}\right)+\left(C_{V}-C_{A}\right)
\left(1-\frac{T_{e}}{E_{\nu}}\right)^{2}-
C_{V}\frac{m_{e}T_{e}}{E_{\nu}^{2}}\right]
$$\begin{equation} 
+\langle r_{\nu}^{2}\rangle^{2}\frac{\pi\alpha^{2}_{\rm e.m.}m_{e}}{9}
\left[1+\left(1-\frac{T_{e}}{E_{\nu}}\right)^{2}-
\frac{m_{e}T_{e}}{E_{\nu}^{2}}\right]
\, .\label{emdiffcross}\end{equation}
The number of observed events as function of the e.m.\ form factors is given
by:
\begin{equation} 
N(\mu_{\nu},\langle r_{\nu}^{2}\rangle)=N_{0}\left[1+
 \frac{\langle\sigma^{M}\rangle}{\langle\sigma^{\rm std}\rangle}\mu_{\nu}^{2}
+\frac{\langle\sigma^{R1}\rangle}{\langle\sigma^{\rm std}\rangle}
\langle r_{\nu}^{2}\rangle+
\frac{\langle\sigma^{R2}\rangle}{\langle\sigma^{\rm std}\rangle}
\langle r_{\nu}^{2}\rangle^{2}\right]
\, ,\label{Nem}\end{equation}
where $N_{0}$ is the standard expectation, and $\langle\sigma^{M}\rangle$, 
$\langle\sigma^{R1}\rangle$, and $\langle\sigma^{R2}\rangle$ are the partial
cross section in Eq.~(\ref{emdiffcross}) integrated on the electron recoil
energy [including the corrections due to the finite resolution of the detector,
according to Eq.~(\ref{crossec})] and folded with the source spectrum. From
Eqs.~(\ref{Nem}) and~(\ref{Limit}) we obtain the equation for 90\% sensitivity
bound in the plane $(\langle r_{\nu}^{2}\rangle,\mu_{\nu})$ for a null result:
\begin{equation} 
\mu_{\nu}=\left[\pm\eta_{0}-\eta_{1}\langle r_{\nu}^{2}\rangle-
\eta_{2}\langle r_{\nu}^{2}\rangle^{2}\right]^{1/2}
\, ,\label{emlimit}\end{equation}
where $\eta_{0}=\langle\sigma^{\rm std}\rangle\varepsilon_{90}/\langle
\sigma^{M}\rangle$ and $\eta_{1,2}=\langle\sigma^{R1,2}\rangle/\langle
\sigma^{M}\rangle$. In Tab.~\ref{tab2} we report our calculation of the
coefficients $\eta_{0}$, $\eta_{1}$, and $\eta_{2}$ for the cases of $^{51}$Cr
and $^{90}$Sr source experiments, and for the MUNU experiment, where
$\mu_{\nu}$ is measured in units of $10^{-10}\mu_{B}$ and $\langle
r_{\nu}^{2}\rangle$ in units of $10^{-32}$~cm$^{2}$.

\begin{figure}
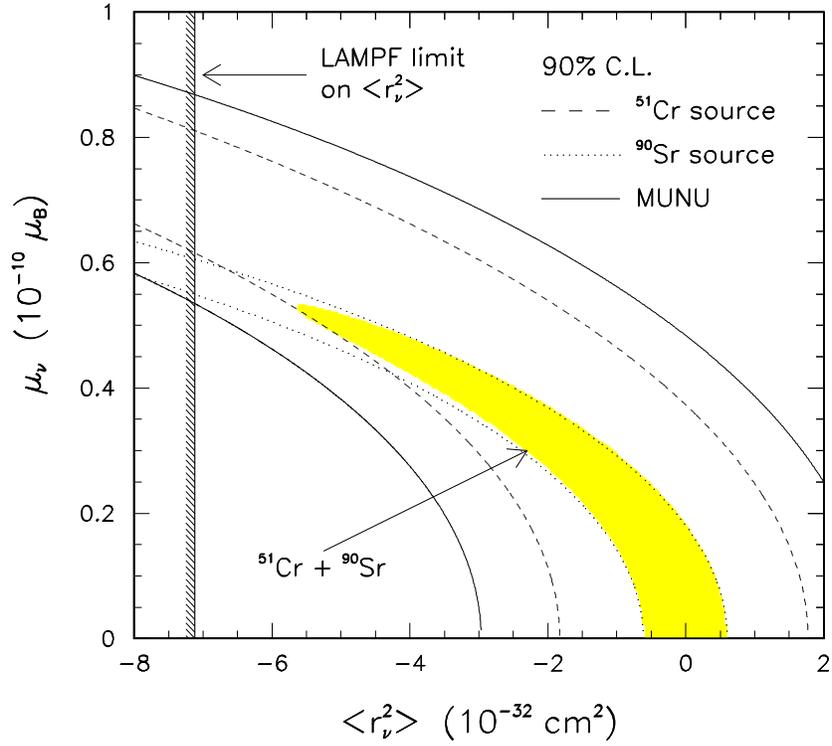

\caption{Prospective 90\% C.L.\ sensitivity contours in the neutrino e.m.\ 
	 form factors plane for the BOREXINO source experiments. Dashed line: 
	 $^{51}$Cr neutrinos; dotted line: $^{90}$Sr antineutrinos; gray area: 
	 combined Sr+Cr experiments. The 90\% C.L.\ LAMPF limit on 
	 $\langle r_{\nu_{e}}^{2} \rangle$ (shaded line) and the expected 
	 90\% C.L.\ MUNU limit (solid line) are also shown.}
\label{munu}
\end{figure}

	In Fig.~\ref{munu} we show the 90\% C.L.\  contours for $^{51}$Cr
neutrinos (dashed lines) and $^{90}$Sr antineutrinos (dotted lines). The shaded
line is the limit on $\langle r_{\nu}^{2}\rangle$ set by LAMPF \cite{LAMPF}. If
no difference with the standard expectation were found, combining this limit
with the BOREXINO measurement one can put an upper limit on $\mu_{\nu}$ equal
to $0.8\times 10^{-10}\mu_{B}$ (90\% C.L.\ for 2 d.o.f.) for neutrinos and
$0.6\times 10^{-10}\mu_{B}$ for antineutrinos. Moreover, one can put an upper
limit on $\langle r_{\nu}^{2}\rangle$ equal to $\simeq 2\times
10^{-32}$~cm$^{2}$ for neutrinos and $\simeq 0.5\times 10^{-32}$~cm$^{2}$ for
antineutrinos, close to the value in \cite{Degrassi}. Moreover, assuming that
the magnetic and anapole moment are equal for neutrinos and antineutrinos, it
is possible to perform a combined fit (gray area in Fig.~\ref{munu}). In this
case, one obtains a more stringent bound on the parameters: $-5.5\leq\langle
r_{\nu}^{2}\rangle/10^{-32}$~cm$^{2}\leq  0.5$ and $\mu_{\nu}\leq 0.55\times
10^{-10}\mu_{B}$.

	In Fig.~\ref{munu} we show for comparison the zone explorable by the
MUNU experiment after one year of operation (solid line). For  $\langle
r_{\nu}^{2}\rangle$ unconstrained, MUNU can put a limit $\mu_{\nu}\leq
0.85\times 10^{-10}\mu_{B}$. For $\langle r_{\nu}^{2}\rangle=0$, we obtain a
limit $\mu_{\nu}\leq 0.42\times 10^{-10}\mu_{B}$ (90\% C.L.\ for 1
d.o.f.),~\footnote{
When only 1 d.o.f. is concerned, $\varepsilon_{90}$ have to be reduced by a
factor 0.767.} 
in relatively good agreement with the MUNU collaboration analysis 
\cite{MUNUexp}.

	From Fig.~\ref{munu} we see that both the $^{51}$Cr and the $^{90}$Sr
limits are more stringent then those expected in MUNU as a result of higher
statistics, of the smaller flux uncertainties, and of the lower energy
threshold of BOREXINO. In particular, the $^{90}$Sr $\bar{\nu}$ experiment is
the most sensitive. For $\langle r_{\nu}^{2}\rangle=0$, the $^{90}$Sr limit on
$\mu_{\nu}$ is $0.16 \times 10^{-10}\mu_{B}$ (90\% C.L.\ for 1 d.o.f.), about
three times better than MUNU. This  limit depends weakly on assumptions about
the source activity and exposure time. For example, with 2.5~MCi activity and 3
months of exposure, this limit is  raised to $0.21\times 10^{-10} \mu_{B}$ ---
about a factor two better than MUNU. In fact, for $\langle
r_{\nu}^{2}\rangle=0$, the limit value of $\mu_{\nu}$ depends on the square
root of $\varepsilon_{90}$ [see  Eq.~(\ref{emlimit})]. This make this
experiment very appropriate for magnetic moment searches.

        \section{Implications for vector and axial couplings}

	The low energy $\nu$-$e$ neutral current interaction is usually
parameterized by the following effective four-fermion Hamiltonian:
\begin{equation}
{\cal H}_{\rm int}^{\nu e}=-\frac{G_F}{\sqrt{2}}\left[
\bar{\psi}_{\nu}\gamma^{\alpha}\left(1-\gamma^{5}\right)\psi_{\nu}
\right]\left[\bar{\psi}_{e}\gamma_{\alpha}
\left(g^{\nu e}_{V}-g^{\nu e}_{A}\gamma^{5}\right)\psi_{e}\right]
\, ,\label{Hint}\end{equation}
where $\psi_{\nu}$ and $\psi_{e}$ are the neutrino and electron fields and
$g^{\nu e}_{V,A}$ are the vector and axial coupling of the neutrino current to
the electron current. When also charge current interactions are involved, as in
the case of $\nu_{e}$-$e$ scattering, $g^{\nu e}_{V,A}\to g^{\nu e}_{V,A}+1$.

	The Standard Model of electroweak interactions states that $g^{\nu
e}_{V}= 2\sin^{2}\theta_{W}-\frac{1}{2}=(-0.038$) and $g^{\nu e}_{A}=-1/2$,
apart from small radiative corrections. Moreover, in the Standard Model 
$g^{\nu e}_{V,A}= g^{\nu}_{V,A}\cdot g^{e}_{V,A}$, where $g^{\nu}_{V,A}$ 
($g^{e}_{V,A}$) are the couplings of the neutrinos (electrons) to the $Z$
boson. The values of  $g^{\nu}_{V,A}$ are inferred from the ``invisible'' decay
width of the $Z$ boson \cite{LEP,PDB}. Although this gives the value of $g^{\nu
e,{\rm std}}_{V,A}$  with great precision, it does not account for possible
non-standard process occurring in the $\nu$-$e$ scattering. (for example,
exchange of non-standard neutral gauge boson, as proposed in \cite{Valle}). For
this reason, a direct measure of the $g^{\nu e}_{V,A}$ couplings is
interesting.

	At present, the most precise {\em direct} determinations of $g^{\nu
e}_{V,A}$, come from the CHARM~II experiment using $\nu_{\mu}$-$e$ scattering
\cite{CHARM}: $g^{\nu e}_{V}=-0.035\pm 0.017$ and $g^{\nu e}_{A}=-0.503\pm
0.017$ at $1\sigma$, in agreement with Standard Model. In this section, we
investigate the possibility to probe $g^{\nu e}_{V,A}$ using the BOREXINO
calibration experiments. From Eq.~(\ref{Hint}), one obtains the differential
cross section for $\nu_{e}$-$e$ scattering \cite{tHooft} in the form of
Eq.~(\ref{diffcross}) with $C_{V,A}=g^{\nu e}_{V,A}+1$ where $g^{\nu e}_{V,A}$
are now independent variables. Expanding Eq.~(\ref{diffcross}) in term of
$g^{\nu e}_{V}$ and $g^{\nu e}_{A}$, and following the same procedure of the
previous section we obtain $N(g^{\nu e}_{V},g^{\nu e}_{A})=N_{0}f(g^{\nu
e}_{V},g^{\nu e}_{A})$ where $N_{0}$ is the standard expectation and
\begin{equation}
f(g^{\nu e}_{V},g^{\nu e}_{A})=
\xi_{V}(g^{\nu e}_{V}+1)^{2}+\xi_{A}(g^{\nu e}_{A}+1)^{2}+
\xi_{VA}(g^{\nu e}_{V}+1)(g^{\nu e}_{A}+1)
\, .\label{NVA}\end{equation}
Here $\xi_{V}=\langle \sigma(C_{V}=1,C_{A}=0)\rangle/ \langle\sigma^{\rm
std}\rangle$, $\xi_{A}=\langle \sigma(C_{V}=0,C_{A}=1)\rangle/
\langle\sigma^{\rm std}\rangle$, and $\xi_{VA}=\langle \sigma(C_{V}=
1,C_{A}=1)\rangle/\langle\sigma^{\rm std}\rangle-\xi_{V}-\xi_{A}$ [$\xi_{VA}$
have opposite sign for $\bar{\nu}$]. In Tab.~\ref{tab3} we show the value of 
the coefficients $\xi_{V}$, $\xi_{A}$, and $\xi_{VA}$ for the cases of
$^{51}${\rm Cr} and $^{90}${\rm Sr} source experiments, and for the MUNU 
experiment.

\begin{figure}
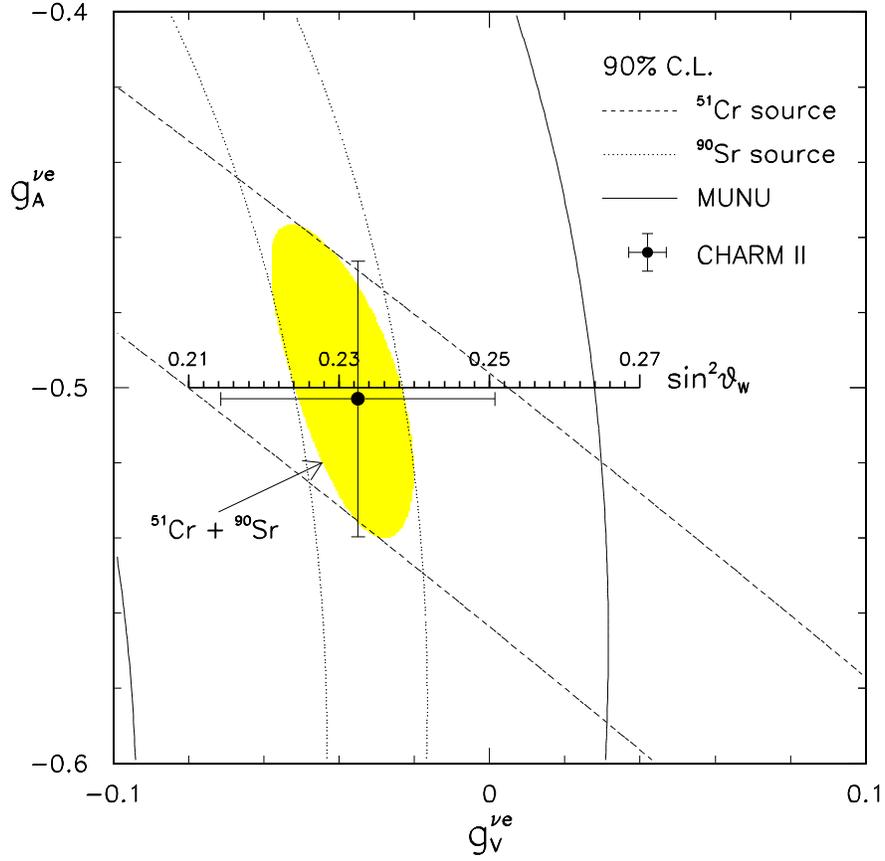

\caption{Prospective 90\% C.L.\ sensitivity contours in the $\nu$-$e$ vector 
	 and axial coupling plane  for the BOREXINO source experiments. Dashed 
	 line: $^{51}$Cr neutrinos; dotted line: $^{90}$Sr antineutrinos; gray 
	 area: combined Sr+Cr experiments. The 90\% C.L.\ data from CHARM~II 
	 data and the expected 90\% C.L.\ MUNU limit (solid line) are also 
	 shown.}
\label{gvga}
\end{figure}

	From Eq.~(\ref{Limit}) we then obtain the 90\% limit in the plane
$(g^{\nu e}_{V},g^{\nu e}_{A})$ for a null result:
\begin{equation} f(g^{\nu e}_{V},g^{\nu e}_{A})=1\pm \varepsilon_{90} \,
.\label{gvgalimit}\end{equation}
This limit is shown in Fig.~\ref{gvga} for the $^{51}$Cr experiment (dashed
line), the $^{90}$Sr experiment (dotted line), and their combination (gray
area). Also shown are the CHARM~II results and the zone explorable by MUNU
(solid line). In BOREXINO, a significative improvement in the measure of
$g^{\nu e}_{V}$ appears possible, whilst the constraints on $g^{\nu e}_{A}$ are
similar to CHARM~II. In particular, we obtain $-0.056\leq g^{\nu e}_{V}\leq
-0.020$ ($0.222\leq\sin^{2}\theta_{W}\leq 0.240$) and $-0.54\leq g^{\nu e}_{A}
\leq -0.46$ at the 90\% C.L. (2 d.o.f.). The $\sin^{2}\theta_{W}$ is thus
measured with a precision of $\sim 4$\% --- a factor two better than CHARM~II.
Fixing $g^{\nu e}_{A}$ to $-1/2$, we obtain a more stringent constraint on the
Weinberg angle: $0.226\leq\sin^2\theta_{W}\leq 0.236$ (90\% C.L.\ for 1 
d.o.f.), corresponding to a precision of $\sim 2.5$\%.

	From Fig.~\ref{gvga} we can see that the BOREXINO constraints are more
stringent than MUNU as a result of higher statistics attainable and of the
combination of $\nu$ and $\bar{\nu}$ signal.

	Finally, we stress that a comparison of scattering experiment of the
kind $\nu_{e}$-$e$ (BOREXINO) and $\nu_{\mu}$-$e$ (CHARM~II) is a useful check
of the universality of weak interactions at low energies.

        \section{Conclusions}

	In this paper we have explored the possibility to search for
non-standard neutrino properties with the BOREXINO Cr and Sr source
experiments. In particular, we have considered  a) Neutrino oscillations; b)
Non-zero electron neutrino e.m.\ form factors $\mu_{\nu}$ and $\langle
r^{2}_{\nu}\rangle$; c) Non-standard $\nu$-$e$ vector and axial couplings. In
case a) we find that, in the channel $\nu_{e}\rightarrow\nu_{s}$, BOREXINO can
extend the oscillation parameter limits for $\delta m^{2}\geq 3$~eV$^{2}$ and  
$0.04\leq\sin^{2}2\theta\leq 0.1$. In case b) BOREXINO can reach a sensitivity
to the magnetic moment equal to $0.8\times 10^{-10}\mu_{B}$ for neutrinos and
$0.6\times 10^{-10}\mu_{B}$ for antineutrinos. In additions, assuming that the
e.m.\ form factors are equal for $\nu$ and $\bar{\nu}$, this limit can be
improved ($\mu_{\nu}\leq 0.5\times 10^{-10}\mu_{B}$) and a limit of
$-5.5\leq\langle r^{2}_{\nu}\rangle/10^{-32}{\rm cm}^{2}\leq 0.5$ can be  put
on the anapole moment --- the strongest limit at present. In the  hypothesis
that $\langle r^{2}_{\nu}\rangle=0$, the  $^{90}$Sr experiment alone can put a
limit  $\mu_{\nu}\leq 0.16\times 10^{-10}\mu_{B}$, improving the MUNU
sensitivity by a factor three. In case c), BOREXINO can reduce the present
uncertainty on the {\em direct} measure of the $g^{\nu e}_{V}$ coupling by a 
factor 2, and can check for the universality of the $\nu$-$e$ interactions  at
low energies. Fixing $g^{\nu e}_{A}$ to $1/2$, the Weinberg angle $\sin^{2}
\theta_{W}$ can be measured with an accuracy of $\pm 2.5$\%.

        \section{Acknowledgments}

	We thank E.\ Lisi for useful suggestions and for careful reading of 
the manuscript, and J.N.\ Bahcall, V.V.\ Sinev, and J.W.F.\ Valle for
suggestions. The work of A.I.\ and G.S.\ was supported in part by Ministero
della Ricerca Scientifica (Dottorato di Ricerca) and in part by INFN. The work
of D.M.\ was supported by INFN.

	\appendix
        \section*{Calculation of the function {\lowercase {\em g}}}

	In this Appendix we give the analytical expression for the function $g$
of Eq.~(\ref{gdraft}). In polar coordinates (choosing the $z$ axis as the FV
center-to-source direction), Eq.~(\ref{gdraft}) reads
\begin{equation}
g=\frac{3}{2}\frac{1}{F(R/D)}\frac{D^{2}}{R^{3}}\int_{0}^{R}r^{2}dr\,
\int_{0}^{\pi}\sin\varphi d\varphi\,
\frac{1}{\delta_{x}^{2}}\cos\frac{\delta m^{2}}{2E_{\nu}}\delta_{x}
\, ,\label{gdraft1}\end{equation}
with $\delta^{2}_{x}=r^{2}+D^{2}-2rD\cos\varphi$. With the substitutions
$x=r/D$ and $y=\beta\delta_{x}/D$ (with $\beta=\delta m^{2}D/2E_{\nu}$) one has
\begin{eqnarray}
g&=&\frac{3}{2}\frac{1}{h^{3}F(h)}\int_{0}^{h} xdx\,
\int_{\beta(1-x)}^{\beta(1+x)}dy\,\frac{\cos y}{y}\nonumber \\
 &=&\frac{3}{2}\frac{1}{h^{3}F(h)}\int_{0}^{h} xdx\,\left[
{\rm Ci}\left(\beta(1-x)\right)-{\rm Ci}\left(\beta(1+x)\right)\right]
\, ,\label{gdraft2}\end{eqnarray}
where $h=R/D$ and the function ${\rm Ci}(z)$ is the integral cosine:
\begin{equation}
{\rm Ci}(z)=-\int_{z}^{\infty }dq\,\frac{\cos q}{q}
\, .\label{Ci}\end{equation}
The integral in Eq.~(\ref{gdraft2}) can be easily evaluated with the help of
the following expressions:
\begin{eqnarray} 
\int dz\, {\rm Ci}(z)  &=& z{\rm Ci}(z)-\sin z\, ,\nonumber\\ 
\int dz\, z {\rm Ci}(z) &=& \frac{z^{2}}{2}{\rm Ci}(z)-
\frac{1}{2}(\cos z+z \sin z) 
\, .\label{Cint}\end{eqnarray}
The final result can be cast in the following form:
\begin{equation}
g(h,\beta) =\frac{3}{4}
\frac{G\left((1+h)\beta,\beta\right)-G\left((1-h)\beta,\beta\right)}
{h^{3}\beta^{2}F(h)}
\, ,\label{g}\end{equation}
where
\begin{equation}
G(z,\beta) =\left[z {\rm Ci}(z)-\sin z\right]
\left(z-2\beta\right)-\cos z
\, .\label{G}\end{equation}




\newpage
\begin{table}\caption{\label{tab1}
Time of exposure ($t_{\rm ex}$), expected background events ($N_{B}$), 
standard signal events ($N_{0}$),  $1\sigma$ uncertainty of $N_{0}$ 
($\delta_{N_{0}}$), and 90\% C.L.\ (2 d.o.f.) relative accuracy for the 
$\nu$-$e$ scattering, $\bar{\nu}$-$e$ scattering, inverse $\beta$-decay, 
and MUNU  experiments.}
\begin{tabular}{lccrrc}
experiment& reaction                 & $t_{\rm ex}$ (days) &$N_{B}$ & 
$N_{0}\pm\delta_{N_{0}}$ &$\varepsilon_{90}$\\ \tableline
$^{51}$Cr & $\nu_{e}$-$e$ scat.      &                  60 &    4380 &   
   4006 $\pm$        100 &$5.4\times 10^{-2}$ \\
$^{90}$Sr & $\bar{\nu}_{e}$-$e$ scat.&                 180 &   17460 &
  25971 $\pm$        333 &$2.8\times 10^{-2}$ \\
$^{90}$Sr & inv. $\beta$-decay       &                 180 & $\sim 5$&
  13278 $\pm$        176 &$2.8 \times 10^{-2}$\\
MUNU      & $\bar{\nu}_{e}$-$e$ scat.&                 365 &    2190 &
   1935 $\pm$        116 &$12.9\times 10^{-2}$
\end{tabular}\end{table}

\vspace{1.4cm}
\begin{table}\caption{\label{tab2}
Coefficients $\eta$ of Eq.~(\ref{emlimit}) for the BOREXINO $^{51}$Cr and 
$^{90}$Sr source experiments, and for MUNU. See the text for details.}
\begin{tabular}{lccc}
source    & $\eta_{0}$       & $\eta_{1}$          & $\eta_{2}$         \\ 
\tableline
$^{51}$Cr & 0.139            & $7.7\times 10^{-2}$ & $6.4\times 10^{-4}$\\
$^{90}$Sr & 0.033            & $5.3\times 10^{-2}$ & $9.1\times 10^{-4}$\\
MUNU      & 0.234            & $8.3\times 10^{-2}$ & $1.4\times 10^{-3}$
\end{tabular}\end{table}

\vspace{1.4cm}
\begin{table}\caption{\label{tab3}
Coefficients $\xi$ of Eq.~(\ref{NVA}) for the BOREXINO $^{51}$Cr and  $^{90}$Sr
source experiments, and for MUNU. See the text for details.}
\begin{tabular}{lccr}
source    & $\xi_{V}$ & $\xi_{A}$ & $\xi_{VA}$ \\ \tableline
$^{51}$Cr &     0.441 &     0.825 &      0.799 \\
$^{90}$Sr &     1.392 &     1.849 &   $-$1.558 \\
MUNU      &     1.359 &     1.695 &   $-$1.422 
\end{tabular}\end{table}


\newcommand{\InsertFigure}[2]{\newpage\begin{center}\mbox{%
\epsfig{bbllx=1.4truecm,bblly=1.3truecm,bburx=19.5truecm,bbury=26.5truecm,%
height=21.truecm,figure=#1}}\end{center}\vspace*{-2.3truecm}%
\parbox[t]{\hsize}{\small\baselineskip=0.5truecm\hskip0.5truecm #2}}

\InsertFigure{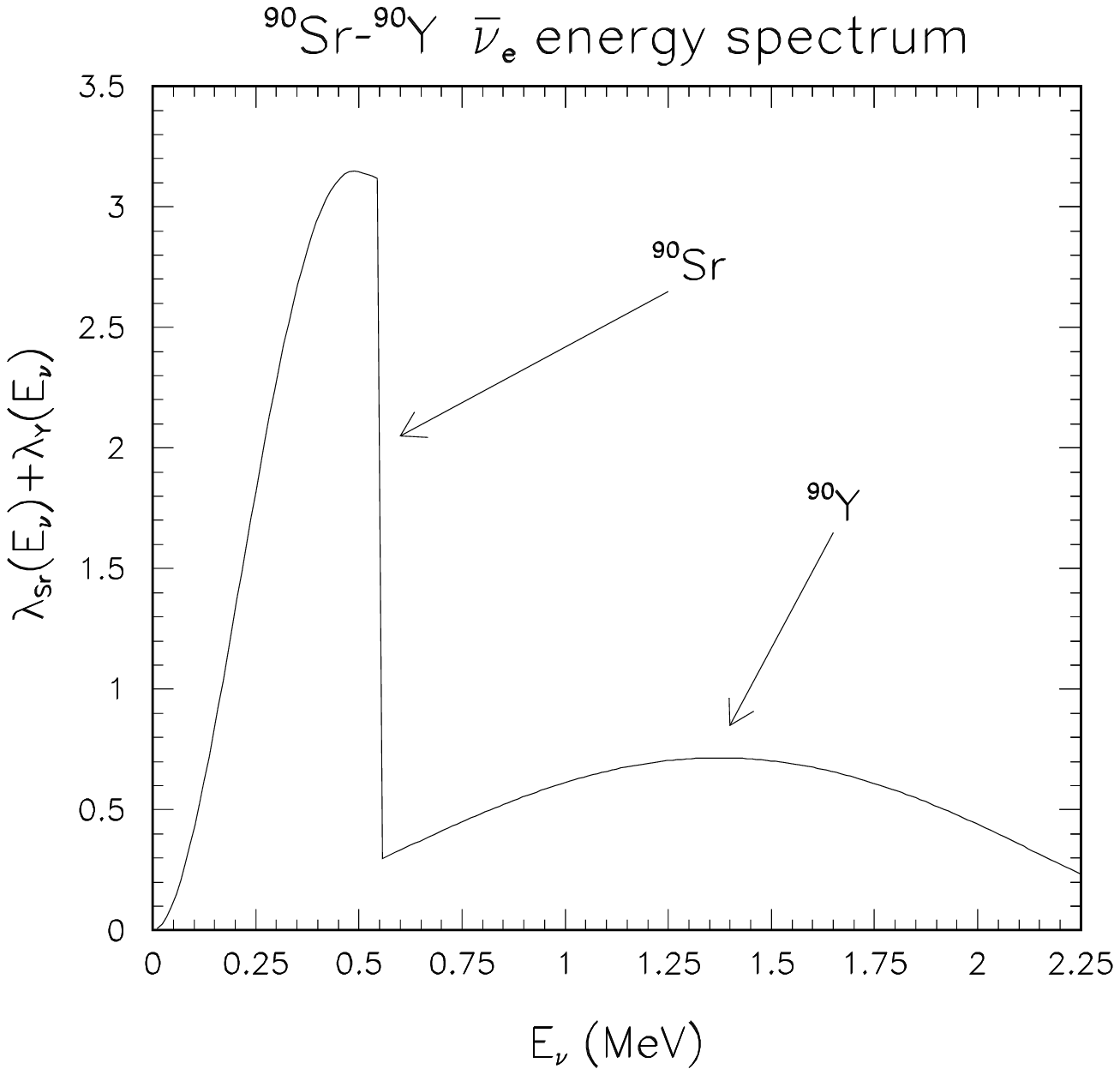}%
{\hfil Fig.~1. Spectrum of the $^{90}$Sr antineutrino source.\hfil}

\InsertFigure{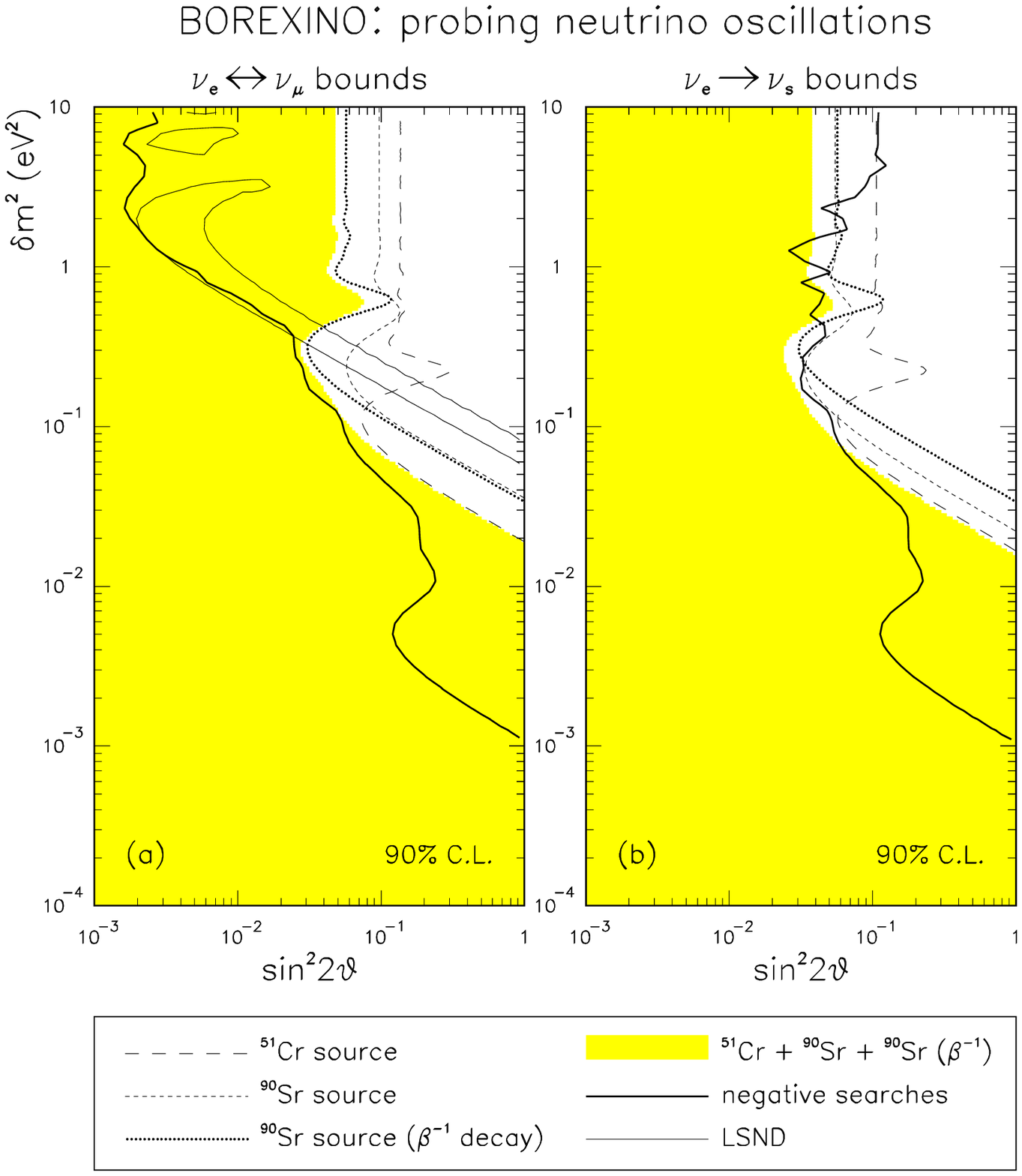}%
{FIG.~2. Prospective 90\% C.L.\ sensitivity contours in the oscillation 
	 parameters plane for the BOREXINO source experiments in the case of 
	 $\nu_{e} \rightarrow\nu_{\mu}$ and $\nu_{e}\rightarrow\nu_{s}$ 
	 transitions [panel (a) and (b), respectively]. Short-dashed line: 
	 $\nu$-$e$ scattering of the $^{90}$Sr antineutrinos; long-dashed line: 
	 $\nu$-$e$ scattering of the $^{51}$Cr neutrinos; dotted line: inverse 
	 $\beta$-decay; gray area: combined Sr+Cr experiments. The 90\% C.L.\ 
	 limits coming from negative evidences of oscillations (solid thick 
	 line) and the LSND allowed area (thin solid line) are also shown.}

\InsertFigure{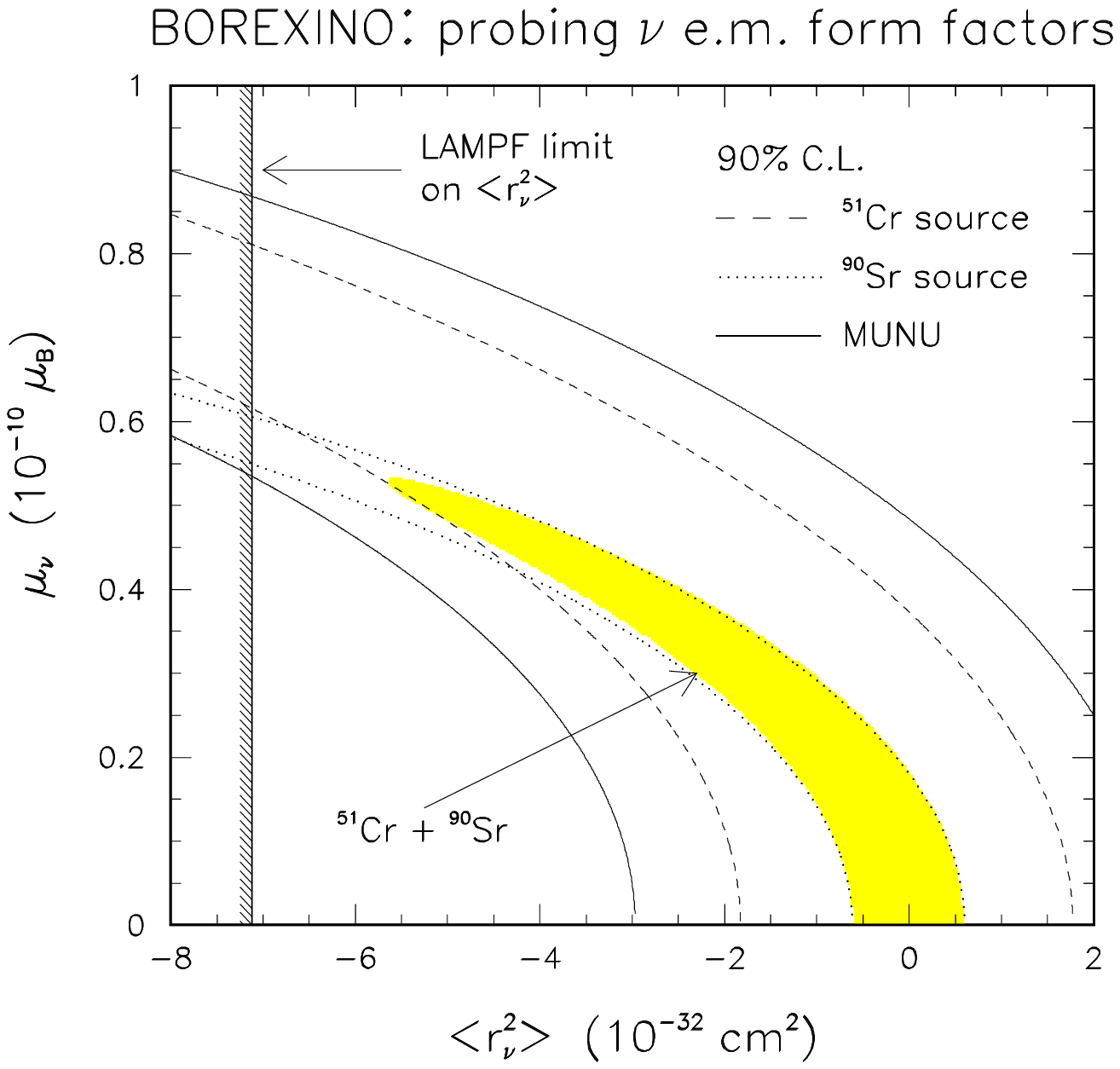}%
{FIG.~3. Prospective 90\% C.L.\ sensitivity contours in the neutrino e.m.\ 
	 form factors plane for the BOREXINO source experiments. Dashed line: 
	 $^{51}$Cr neutrinos; dotted line: $^{90}$Sr antineutrinos; gray area: 
	 combined Sr+Cr experiments. The 90\% C.L.\ LAMPF limit on 
	 $\langle r_{\nu_{e}}^{2} \rangle$ (shaded line) and the expected 
	 90\% C.L.\ MUNU limit (solid line) are also shown.}

\InsertFigure{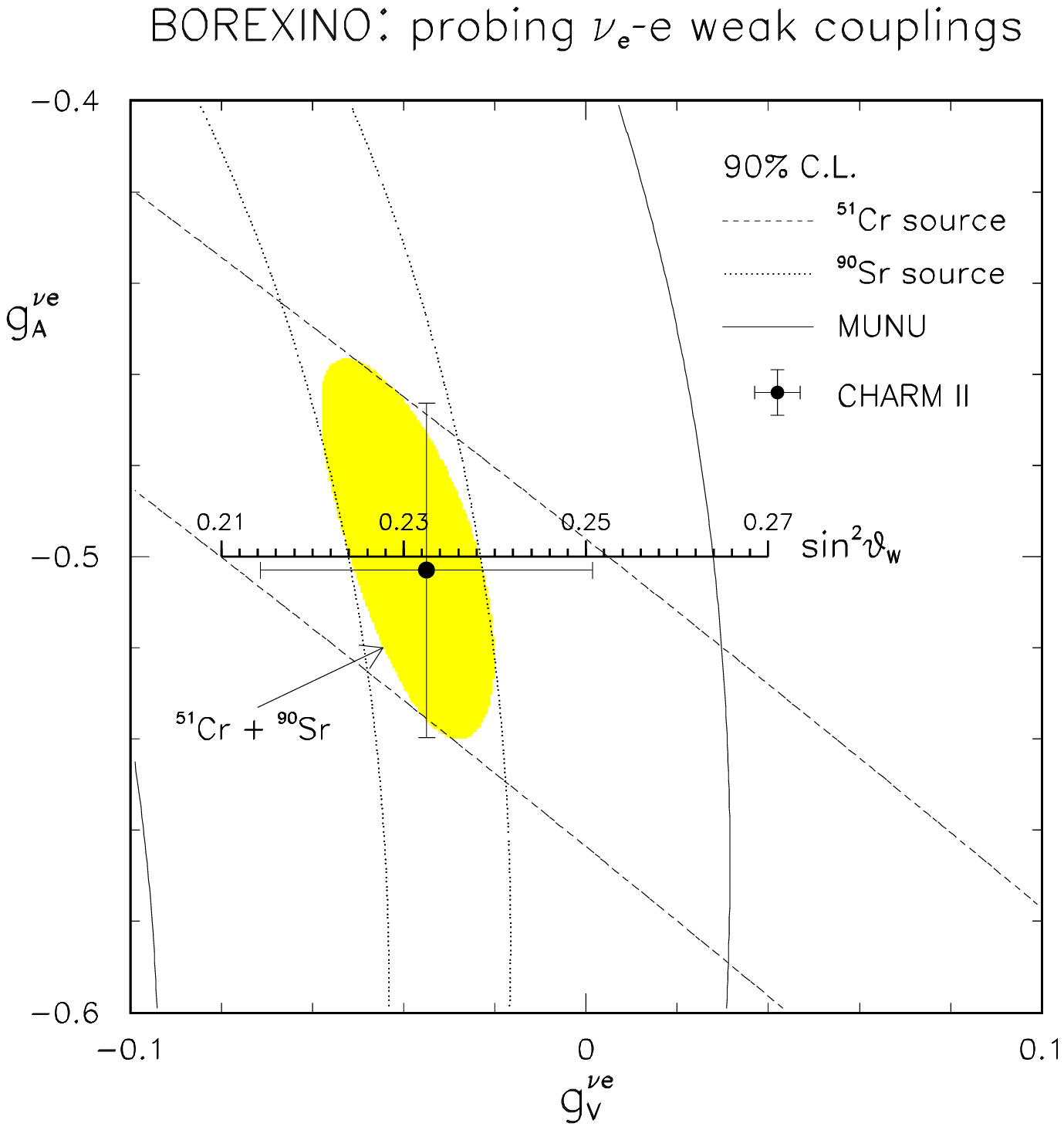}%
{FIG.~4. Prospective 90\% C.L.\ sensitivity contours in the $\nu$-$e$ vector 
	 and axial coupling plane  for the BOREXINO source experiments. Dashed 
	 line: $^{51}$Cr neutrinos; dotted line: $^{90}$Sr antineutrinos; gray 
	 area: combined Sr+Cr experiments. The 90\% C.L.\ data from CHARM~II 
	 data and the expected 90\% C.L.\ MUNU limit (solid line) are also 
	 shown.}

\end{document}